\documentclass[preprint]{aastex}  
\usepackage{emulateapj5}
\usepackage{psfig}

\slugcomment{Draft of \today}
\submitted{Accepted for publication in the Astronomical Journal}

\shortauthors{Strickland \etal}
\shorttitle{New insights
into the nature of starburst-driven superwinds}

\newcommand{\eg}{{\rm e.g.\ }}

\newcommand{\ie}{{\it i.e.\ }}
\newcommand{\cf}{{\rm cf.\ }}
\newcommand{\etal}{{\rm et al.\thinspace}}

\newcommand{\cm}{{\rm\thinspace cm}}
\newcommand{\km}{{\rm\thinspace km}}
\newcommand{\pcc}{\hbox{$\cm^{-3}\,$}}
\newcommand{\s}{{\rm\thinspace s}}
\newcommand{\yr}{{\rm\thinspace yr}}

\newcommand{\ps}{\hbox{\s$^{-1}\,$}}
\newcommand{\pyr}{\hbox{\yr$^{-1}$}}

\newcommand{\kmps}{\hbox{$\km\s^{-1}\,$}}
\newcommand{\pcmsq}{\hbox{$\cm^{-2}\,$} }

\newcommand{\counts}{{\rm\thinspace counts}}
\newcommand{\halpha}{H$\alpha$}

\newcommand{\K}{{\rm K}}
\newcommand{\hi}{H{\sc i}}
\newcommand{\hii}{H{\sc ii}}
\newcommand{\nii}{[N{\sc ii}]}

\newcommand{\nH}{\hbox{$N_{\rm H}$}}
\newcommand{\Mdot}{\hbox{$\dot M$}}

\newcommand{\pc}{{\rm\thinspace pc}}
\newcommand{\kpc}{{\rm\thinspace kpc}}
\newcommand{\Mpc}{{\rm\thinspace Mpc}}
\newcommand{\keV}{{\rm\thinspace keV}}

\newcommand{\Lsol}{\hbox{$\thinspace L_{\sun}$}}
\newcommand{\Msol}{\hbox{$\thinspace M_{\sun}$}}
\newcommand{\Zsol}{\hbox{$\thinspace Z_{\sun}$}}

\begin{document}

\title{Chandra observations of NGC 253: New insights
into the nature of starburst-driven superwinds}

\author{David K. Strickland and Timothy M. Heckman}
\affil{Department of Physics and Astronomy, The Johns Hopkins University,
	3400 North Charles Street, Baltimore, MD 21218 \vspace{2mm}}
\email{dks@pha.jhu.edu, heckman@pha.jhu.edu}

\author{Kimberly A. Weaver}
\affil{NASA/Goddard Space Flight Center, Code 662, Greenbelt, Maryland 20771}
\email{kweaver@cleo.gsfc.nasa.gov}

\and
\author{Michael Dahlem}
\affil{Sterrewacht Leiden, Postbus 9513, 2300 RA Leiden,
	The Netherlands}
\email{mdahlem@strw.leidenuniv.nl}

\begin{abstract}
Arcsecond-resolution X-ray imaging of the nucleus of the nearby starburst
galaxy NGC 253 with {\it Chandra} reveals a well-collimated, strongly
limb-brightened, kiloparsec-scale conical outflow from the central 
starburst region. The outflow is very similar in morphology to the
known \halpha~outflow cone, on scales down to $\lesssim 20 \pc$.
This provides, for the first time, robust evidence that both 
X-ray and \halpha~emission come from low volume filling factor regions
of interaction between the fast energetic wind of SN-ejecta
and the denser ambient interstellar medium (ISM), 
and not from the wind fluid itself.
We provide estimates of the (observationally
and theoretically important) filling factor of the
X-ray emitting gas, of between $\sim 4$ and $40$ per cent, consistent with
an upper limit of $\sim 40$ per cent based directly on the observed 
limb-brightened morphology
of the outflow. Only $\lesssim 20$ per cent of the observed X-ray emission
can come from the volume-filling, metal-enriched, wind fluid itself.
Spatially-resolved spectroscopy of the soft diffuse thermal X-ray emission
reveals that the predominant source of spectral variation along the
outflow cones is due to strong variation in the absorption, on scales
of $\sim 60 \pc$, there being little change in the characteristic
temperature of the emission. 
We show that these observations are easily explained by, and fully consistent
with, the standard model of a superwind driven by a starburst of
NGC 253's observed power. 
If these results are typical of all starburst-driven winds,
then we do not directly see all
the energy and gas (in particular the hot metal-enriched gas) transported
out of galaxies by superwinds, even in X-ray emission. 

\end{abstract}


\section{Introduction}
\label{sec:introduction}

Edge-on local starburst galaxies show unambiguous evidence for
$\sim 10$ kiloparsec-scale, weakly-collimated, bipolar 
outflows \citep{heck93}. 
Current theory holds that these galactic superwinds are 
powered by the collective mechanical power of large numbers
of Type II supernovae (SNe) and stellar winds, that result from the
large population of massive stars formed in the starburst 
(\citet{cc}, henceforth CC).
If this mechanical energy is efficiently thermalized in the
starburst region (\ie converted back into the thermal energy of a hot
gas by shocks, as SNe and stellar winds collide), then
a pressure-driven outflow from the galaxy results. The hot gas
blows out of the host galaxy's ISM along the minor axis, 
forming the outflows seen in non-thermal radio emission, 
optical emission lines, and soft thermal X-ray emission 
in the halos of local starbursts.

Superwinds are of cosmological interest as they transport large amounts
of gas, in particular newly synthesized heavy elements, and
energy, into the inter-galactic medium (IGM). Quantifying this
mass, metal and energy transport in local starburst galaxies 
is essential for understanding the
significance of outflows from star-forming galaxies
integrated over the history of the Universe.
However, even the basic physical properties of local superwinds
such as mass outflow rates, energy content, abundances and
kinematics are uncertain.

Measuring the physical properties of the hot gas driving these
outflows is of crucial importance for several simple
reasons. Firstly, the hot gas efficiently transports the
majority of the energy of the outflow (\cf \citet{ss2000}). 
Secondly, the SN-heated gas is thought to
contain the majority of the newly synthesized metals. Finally,
this energetic gas ultimately controls the ejection of mass from the galaxy
(although the majority of the mass of the outflow may be in 
ambient ISM swept-up by the wind, this gas is accelerated
to high velocity by the hot, fast, wind fluid, \cf CC).

\begin{figure*}[!th]
\centerline{
  \psfig{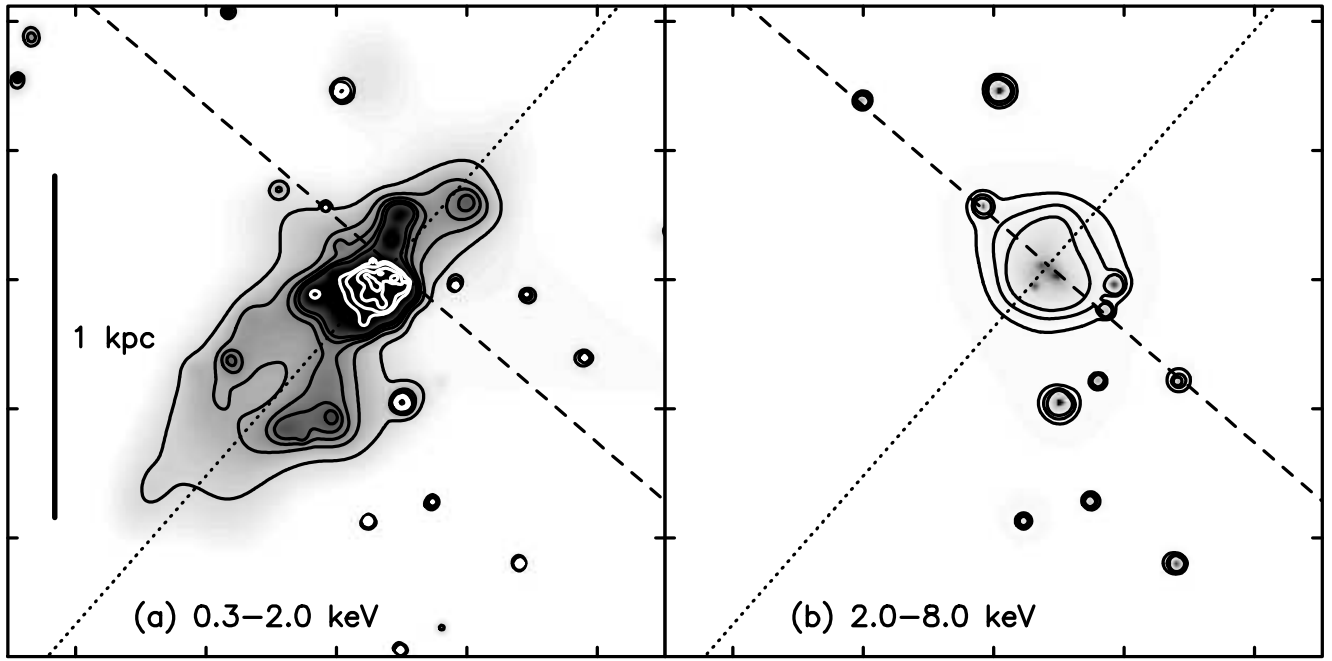}}
\caption{
   (a) {\it Chandra} ACIS S3 soft X-ray (0.3 -- 2.0 \keV) 
   image of the center of NGC 253. Tick marks correspond to $30\arcsec$.
   The image, shown on a square-root intensity scale, has been 
   adaptively smoothed and background subtracted.
   Contour levels shown in black start at 
   $2.9 \times 10^{-5}  \counts \ps {\rm arcsec}^{-2}$, and increase
   in steps of $3.2 \times 10^{-5} \counts \ps {\rm arcsec}^{-2}$. 
   White contours highlight the complex high surface brightness emission
   within the starburst region itself, at levels of
   (4.8, 6.5, 13.0, 26.0) $\times 10^{-4} \counts \ps {\rm arcsec}^{-2}$.
   (b) The hard X-ray image (2.0 -- 8.0 keV energy band) of the same
   region as in (a), again adaptively smoothed and shown using a square-root
   intensity scale. Contour levels have been chosen to highlight the
   clear {\em lack} of hard X-ray emission along the outflow lobes seen in (a).
   The black contour levels start at 
   $0.7 \times 10^{-5}  \counts \ps {\rm arcsec}^{-2}$ (a factor 4
   lower than the minimum used in (a)), and increase
   in multiples of two in surface brightness up to the minimum surface
   brightness level used in the contours of (a).
   In both panels the dashed and dotted lines
   show the position of the major and minor axes of the galaxy respectively,
   which intersect at the position of the brightest radio source
   (J2000.0, $\alpha$ = 00 47 33.14, $\delta$ = -25 17 17.2), 
   which is presumed to be the nucleus \citep{ua97}. 
}
\label{fig:xray1}
\end{figure*}

In principle, X-ray observations of the thermal
emission from hot gas in superwinds can be used to measure the
properties of the hot gas. Large amounts of {\it ROSAT} and {\it ASCA} X-ray
data on starbursts already exists 
(see \citet{rps97}; \citet{ptak}; \citet{dwh}, among others).
However, the origin of the
soft X-ray emission seen in local superwinds is still ambiguous.
Plausible alternatives are:
\begin{enumerate}
\item A hot, adiabatically-expanding,
  volume filling fluid of stellar wind and SN-ejecta material, 
  \ie the metal-enriched, energy containing 
  ``wind'' itself 
  (\eg \citet{fab88}; \citet{breg95}).
\item A strongly ``mass-loaded'' wind, where the gas density 
  of the hot wind has been significantly increased by 
  efficiently mixing in cooler ambient gas (\eg \citet{suchkov96}),
  increasing the density 
  throughout most of the wind's volume. Mass-loading probably occurs
  primarily within the starburst region.
\item Emission from low 
  filling factor regions of interaction between the hot, high
  velocity, wind and clumps or clouds of ambient ISM overrun by the
  wind. The X-ray emitting gas in and/or around the clumps
  may be heated by shocks (\cf CC), 
  by localized hydrodynamical mixing of hot wind
  and cooler cloud material, or by cloud evaporation due to thermal 
  conduction. 
\end{enumerate}


In the first two options X-ray observations probe the wind itself,
\ie the gas filling the majority of the volume, containing the
majority of the energy, and the majority of the metals
(although heavily diluted in case 2 by ambient ISM). In the third
case X-ray observations probe low filling factor material that
contains little energy, mass or metal-enriched gas. It is therefore
essential to know the origin of the X-ray emission, both to correctly derive
the properties of the X-ray emitting gas, and by extension 
those of the superwind as a whole.

In this Paper, we demonstrate the first direct observational
evidence that soft thermal X-ray emission from starburst-driven superwinds
arises in regions of interaction between denser ambient ISM and
the hot, fast, wind fluid (case 3 above), based on {\it Chandra} ACIS
X-ray observations of the nearby starburst galaxy NGC 253.

NGC 253 is one of the two brightest, closest\footnote{We assume the
distance of $D=2.6\pm{0.7} \Mpc$ to NGC 253, given by \citet{puche88}, 
for all calculations.} and best studied
starburst galaxies (the other is M82). The starburst in NGC 253
has a bolometric luminosity of about $2 \times 10^{10} \Lsol$,
and so is situated right at the ``knee'' of the starburst
luminosity function \citep{soifer87}. 
Thus, starbursts with luminosities similar to
NGC 253 dominate the starburst ``emissivity'' of the local universe.
NGC 253 is seen almost edge-on, ideally oriented for studying the
superwind as it flows out into the galaxy halo.

While both of the above attributes are shared by M82, NGC 253 offers
two distinct advantages. Firstly, it lies behind a much smaller
Galactic \hi~column than M82 ($10^{20}$ 
vs. $4 \times 10^{20} \pcmsq$ respectively).
Since the superwind phenomenon is most conspicuous in soft X-rays,
which are easily absorbed,
this is a significant advantage. Secondly, NGC 253 (a late-type
$L_{\star}$ spiral) is much more typical of luminous starbursts
than is M82 (a peculiar dwarf amorphous galaxy). Thus, we expect the
lessons learnt from NGC 253 to be more reliably generalizable.

\bigskip
\begin{minipage}[l]{0.46\textwidth}
\centerline{
\psfig{figure=strickland.fig2.eps,%
width={\columnwidth},bbllx=105bp,bblly=230bp,bburx=435bp,bbury=560bp,clip=t}
}
  \medskip
  \noindent\begin{minipage}{0.99\linewidth} 
    \vskip \abovecaptionskip\footnotesize\noindent
    Fig. 2. --- Logarithmically scaled continuum-subtracted 
    \halpha~image of the same region of NGC 253 as shown in Fig.~1.
    The 0.3--2.0 keV X-ray emission is shown as white contours.
    There is a clear similarity between the soft X-ray and \halpha~emission,
    in particular the southern outflow
    cone is strongly limb-brightened in both soft X-ray and \halpha~emission.
\label{fig:ha1}
   \par\vskip \belowcaptionskip
  \end{minipage}
\end{minipage}

\section{Data analysis}
\label{sec:data_analysis}

Guest observer observations of NGC 253, using the 
AXAF CCD Imaging Spectrometer (ACIS\footnote{{\it Chandra} Observatory Guide 
\url{http://asc.harvard.edu/udocs/docs/docs.html}, 
section ``Observatory Guide'',``ACIS''}) 
on board the {\it Chandra}
X-ray Observatory, were obtained on 1999 December 16, 
with the nuclear starburst region placed on the back-illuminated
CCD chip S3.

Here we report on the soft X-ray properties of the outflow within
the central kiloparsec-scale region of NGC 253, and defer further
discussion, and detailed spectroscopic study, of X-ray emission from 
point sources and the larger-scale diffuse X-ray emission  
to forthcoming papers (\citet{whsd2000}; \citet{shwd2000}).

Data reduction and analysis was performed using {\sc Ciao} (v1.1.4)
and {\sc Heasoft} (v5.0.1). The data were reprocessed using the latest
gain correction file available at the time of writing
(acisD1999-09-16gainN0003.fits), screened
to remove ``bad'' events, and time filtered to remove periods of
high background (``flares,'' where the total count rate in any chip deviated
by more than $5\sigma$ from the mean). This resulted in a
remaining exposure length of 12862s on chip S3.

Background subtraction for both imaging and spectral
analysis was performed using the background event files and
software provided by the {\it Chandra} X-ray Center (CXC), scaled
to the exposure time of our observation.

Spectra were extracted using Pulse Invariant (PI) data values, in order to
account for the spatial variations in gain between different
chip nodes. In order to use $\chi^{2}$ as the fit statistic 
the spectra were rebinned to achieve a minimum of
at least 10 counts per bin.
Given the calibration uncertainties in the ACIS BI chip spectral
response (in particular at energies below $E \sim 1$ keV), 
we seek only to use spectral fitting to qualitatively 
characterize the emission, and do not place too much significance
in the numerological results of the fits themselves.

Spectral responses were created using the latest calibration files
available for observations taken with the CCD temperature of
-110 $\degr C$ (the April 2000 release), 
following the example procedures provided
by the CXC\footnote{See the {\sc Ciao} Science Threads: 
\url{http://asc.harvard.edu/ciao/threads/threads.html}}.
We used the spectral response appropriate for the flux-weighted center
of the spatial region from which any spectrum was extracted
(typically a region $\sim 40 \times 100$ raw pixels in size,
see \S~\ref{sec:spectral_structure}).
This does not fully account for the spatial variation in the ACIS
spectral response, but is a good first approximation, given the current
lack of publically available software to construct spectral
responses for spectra from regions larger than the $32 \times 32$
pixel regions over which a single response strictly applies.

The data products initially released
to us by the CXC suffered from processing errors that introduced an
$\sim 9\arcsec$ shift in the absolute coordinate system for the data.
For the purposes of this Paper, we registered the {\it Chandra}
data against ground-based R-band and continuum-subtracted \halpha~data,
based on 9 secure optical cross-identifications of X-ray sources 
seen in the {\it Chandra} observation. This is accurate to $\sim 0.9\arcsec$
(root mean square).

\section{General features of the X-ray data}
\label{sec:results}

\subsection{Soft X-ray emission}

Many different sources of X-ray emission are visible in 
the {\it Chandra} soft X-ray ($0.3$ -- $2.0 \keV$  energy band) image
of NGC 253 (Fig.~1a).
Emerging from the nuclear starburst region, bright diffuse soft X-ray emission
can be seen extending away from the plane of the galaxy 
to the south east and north west for a distance of
$\sim 1\farcm2$ and $0\farcm6$ respectively 
($\sim 900 \pc$ and $\sim 450 \pc$, assuming $D = 2.6 \Mpc$).
This is the innermost part of the starburst-driven superwind, which has
been traced with {\it ROSAT} out to $\sim 9 \kpc$ from the nucleus
\citep{dwh}. 
The southern lobe seen by {\it Chandra} 
is well described as a limb-brightened conical
outflow, of opening angle $\theta \sim 26\degr$, with a truncated base of 
diameter $\sim 24\arcsec$, similar in size to that of the starburst region 
itself. The outflow cone is also seen in \halpha~images,
and the positions and size of the X-ray and \halpha~outflows 
match up almost exactly (see Fig.~2).
Beyond a distance of $\sim 600 \pc$ from the nucleus, 
the X-ray emission becomes
too faint to trace the limb-brightening further, and the \halpha~morphology
at this point becomes confused due to emission from what are
known to be bright \hii~regions along the 
spiral arm seen in the background \citep{mhvb87}.
We analyze the spatial structure of the limb-brightened region of the 
southern outflow lobe in more detail in \S~\ref{sec:outflow_structure}.

Due to NGC 253's inclination, the northern outflow lobe is seen 
through the disk of the galaxy.
Strong absorption (see \S~\ref{sec:spectral_structure}) 
makes the inner region of
the northern outflow less obviously conical, and
it can not be traced as far as the southern outflow cone in the
{\it Chandra} data.

Many X-ray point sources are also clearly visible, 
presumably high mass X-ray binaries or supernova remnants. 

The peaks in the X-ray surface brightness at the south-eastern-most
extent of the limb-brightened region are resolved structures, and
are not point-like.
The effective spatial resolution of the data within the region shown
in Fig.~1,  measured by fitting 2-dimensional Gaussian models to
the brightest point sources, is $\sim 0\farcs9$. As the pixel size
of the ACIS instrument is $0\farcs492$, point sources are
easily distinguished from diffuse emission in unsmoothed, unbinned, images,
as most of the counts of a true point source can be found within
a few adjacent pixels.

\subsection{Hard X-ray emission}

Hard X-ray emission from NGC 253, in the energy band 2.0 -- 8.0 keV,
is dominated by point source emission (see Fig.~1b). Several
point-like sources are visible in the central starburst region 
(at least three within the central $8\arcsec$ [$\sim 100$ pc]
radius), although these are {\em not} the brightest X-ray sources in
NGC 253. 

The nucleus is surrounded by what appears to be extended
hard X-ray emission, roughly spherical but somewhat elongated 
preferentially along the {\em major axis} of the galaxy, 
which we can trace out to a radius of $\sim 8\arcsec$
(100 pc) with confidence (in that this size is apparent to the eye in the 
raw, unsmoothed, data). At lower surface brightness levels the
increasing apparent size and sphericity 
of this diffuse hard X-ray emission may be an
artefact of the adaptive smoothing algorithm used.
The origin and properties of this emission will be discussed
in \citet{whsd2000}.

There is no significant hard X-ray emission from the the southern
outflow cone, which is so clearly seen in the soft X-ray band.
We shall discuss what this implies for the presence of very hot
gas ($kT \ge 2$ keV) in the outflow in \S~\ref{sec:spectral_structure} \& 
\ref{sec:outflow_structure}.


\section{Spectral structure within the outflow}
\label{sec:spectral_structure}

We shall not attempt a rigorous, quantitative, spectroscopic study of the
outflow in this Paper given the uncertainties in the spectral
calibration of the ACIS instrument.
We believe that the data can support 
two {\em robust} conclusions:
\begin{enumerate}
\item The primary cause of spectral variation along, and between, 
  the northern and southern outflow cones is due to variations in the
  absorption column, and not due to significant temperature variations
  along the outflow.
\item Features in the ACIS spectra of the diffuse emission show
  clear evidence for X-ray line emission from highly ionized $\alpha$-elements
  (\eg Mg and Si, see Fig.~3). 
  This clearly demonstrates the thermal nature of the X-ray
  emission. 
\end{enumerate}

\setcounter{figure}{2}

\begin{figure*}[!th]
 \centerline{
 \psfig{figure=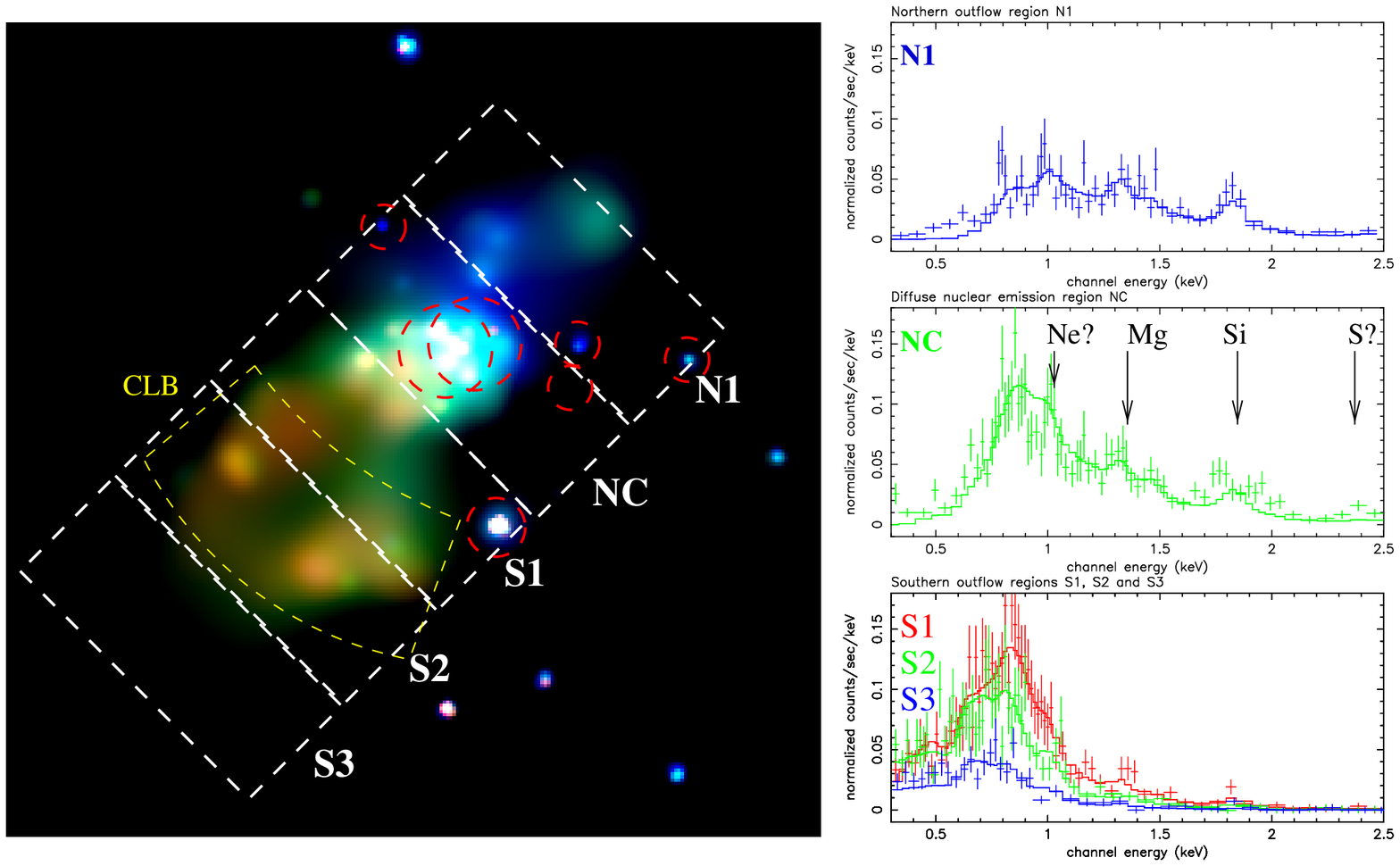,%
 width={2\columnwidth}
 }
 }
 \caption{Three color, logarithmically-scaled, 
  soft X-ray image of the central 
  $2\arcmin \times 2\arcmin$ ($\sim 1.5 \times 1.5 \kpc^{2}$)
  region, showing the spectral variation along the outflow lobes.
  Emission between $0.3 \le E (\keV) < 0.6$ is shown in red,
  emission between $0.6 \le E (\keV) < 1.1$ in green,
  and emission between $1.1 \le E (\keV) < 2.0$ in blue. 
  The south-eastern outflow cone, on the near side of the disk,
  becomes progressively ``soft'' as it rises above the absorbing \hi~gas in
  NGC 253's disk and starburst region. In contrast, only the
  highest energy X-ray photons from the north-western
  lobe of the outflow make it through the disk without absorption.
  The white dashed boxes show regions from which spectra
  of the diffuse emission, shown in the panels on the right,
  were extracted. The yellow dashed line outlines the region
  most clearly limb brightened in both X-ray and \halpha~images.
  Dashed red circles show regions of emission
  from point sources that were excluded in extracting the
  diffuse spectra. The spectra of the nuclear region and the northern outflow
  clearly show significant absorption at photon energies $E < 1 \keV$.
  }
 \label{fig:n253_fig_rgb}
\end{figure*}

\subsection{Simple spectral fitting}
\label{sec:spectra}

To roughly characterize the diffuse X-ray emission from the inner
kiloparsec of the superwind, \ie 
an approximate temperature and amount of foreground absorption,
we divided the outflow lobes and nuclear starburst region into
the set of regions shown in Fig.~3. From each region we then
extracted the spectrum of the diffuse emission, having excluded
any detected point sources, and used a simple one-temperature
hot plasma model to fit the spectrum. 

The region of the southern outflow lobe most clearly limb
brightened, in both the X-ray and \halpha~data, falls within
our spectral regions S1 and S2. As this limb-brightened
region if of particular interest (and is the focus of 
\S~\ref{sec:outflow_structure}), we also extracted and fitted
the spectrum (Fig.~4) 
of this region (henceforth called the CLB region).

Only the data within the
energy range between $E = 0.3$ to $3.0$ keV was used in the spectral fitting,
given that the 2.0 -- 8.0 keV ACIS count rates for all the regions
(except the starburst region NC and the northern lobe region N1) are
essentially zero (see Table~1). The lack of hard X-ray emission
in the southern outflow lobe is graphically demonstrated in
Figs.~\ref{fig:xray1} \& 4.

To begin with, we fitted a variable abundance, single temperature {\sc Mekal}
hot plasma model to the combined spectra from the southern outflow cone,
(regions S1, S2 \& S3 in Fig.~3) where a hardness ratio analysis 
(see \S~\ref{sec:hardness}) already
indicated low absorbing columns and similar spectral shape.  
The abundance ratios (with respect to the Solar value) 
of the $\alpha$-elements were constrained to
be equal to each other, while the abundance ratios of the other elements
were forced to be equal to that of Iron.

{\footnotesize
 \renewcommand{\arraystretch}{1.4}
 \renewcommand{\tabcolsep}{1mm}
 \begin{center}
 TABLE 1
 \vspace{1mm}
 
 {\sc Spectral fits to diffuse X-ray emission regions}
 \vspace{1mm}

\begin{tabular}{lccccr}
\hline
\hline
Region & 
\multicolumn{2}{c}{Count rate$^{a}$} &     
$\nH^{b}$ & 
$kT^{b}$ & 
$\chi^2$/d.o.f \\
            & 0.3--2.0 keV & 2.0--8.0 keV 
	& $10^{20} \pcmsq$ & $\keV$    &                \\
\hline

N1  & $4.81\pm{0.20}$ & $0.64\pm{0.16}$ & 
	$92.2^{+16.6}_{-15.9}$ & $0.59^{+0.12}_{-0.09}$ & 77.5 / 49 \\
NC  & $7.67\pm{0.24}$ & $1.08\pm{0.17}$ & 
	$47.3^{+11.6}_{-11.8}$ & $0.66^{+0.10}_{-0.08}$ & 128.5 / 67 \\
S1  & $7.68\pm{0.25}$ & $0.15\pm{0.16}$ & 
	$5.6^{+3.0}_{-2.2}$ & $0.63^{+0.05}_{-0.06}$ & 42.8 / 57 \\
S2  & $5.74\pm{0.22}$ & $0.07\pm{0.16}$ & 
	$\le 3.4$ & $0.51^{+0.07}_{-0.06}$ & 53.1 / 45 \\
S3  & $2.52\pm{0.15}$ & $0.05\pm{0.16}$ & 
	$\le 5.0$ & $0.46^{+0.13}_{-0.10}$ & 37.8 / 22 \\
CLB & $7.00\pm{0.20}$ & $0.09\pm{0.16}$ & 
	$2.1^{+2.1}_{-1.3}$ & $0.57^{+0.06}_{-0.06}$ & 50.7 / 53 \\
\hline
\end{tabular}
\end{center}
$^{a}$ Background-subtracted ACIS S3
  count rates in units of \mbox{$10^{-2}$ count s$^{-1}$}.\\
$^{b}$ Confidence regions are 90\% confidence
  for three free parameters ($\Delta \chi^{2} = 6.25$).\\}

{\footnotesize
 \renewcommand{\arraystretch}{1.4}
 \renewcommand{\tabcolsep}{1mm}
 \begin{center}
 TABLE 2
 \vspace{1mm}
 
 {\sc Hardness ratios in region of clear limb brightening}
 \vspace{1mm}


\begin{tabular}{lcccr}
\hline
\hline
Region & PA$^{a}$ &
\multicolumn{2}{c}{Count rates$^{b}$} &     
$Q^{c}$ \\ 
           & deg & 0.3 -- 1.0 keV & 1.1 -- 2.0 keV & \\
\hline
Entire CLB & 130--160 & $6.06\pm{0.22}$ & $0.87\pm{0.09}$ & $-0.75\pm{0.04}$ \\
East limb  & 130--137 & $1.58\pm{0.11}$ & $0.21\pm{0.07}$ & $-0.77\pm{0.09}$ \\
Center     & 137--144 & $0.72\pm{0.08}$ & $0.10\pm{0.06}$ & $-0.76\pm{0.16}$ \\
West limb  & 144--160 & $3.65\pm{0.17}$ & $0.55\pm{0.07}$ & $-0.74\pm{0.06}$ \\
\hline
\end{tabular}
\end{center}
$^{a}$ Range of position angle in the
  azimuthal region (see Fig.~6.\\
$^{b}$ Background-subtracted ACIS S3 
  count rates in units of \mbox{$10^{-2}$ count s$^{-1}$}.\\
$^{c}$ Hardness ratio $Q = (H-S)/(H+S)$, where $H$ is the count rate
  in the 1.1 -- 2.0 keV band, and $S$ is the count rate in the 0.3 -- 1.1
  keV band.\\}
\vspace{2mm}

This spectral fit, although of a region of very low and relatively uniform
hydrogen column density (to minimize possible spectral complexity), 
gives strongly sub-solar abundance 
ratios ($Z_{\alpha} \sim 0.30^{+0.14}_{-0.10} \Zsol$, 
$Z_{\rm Fe} \sim 0.13^{+0.04}_{-0.03} \Zsol$, 90\% 
confidence in one interesting parameter). 
This is likely to be a combination of further spectral complexity 
on small scales (emission from multi-phase
gas, \cf \citet{ss2000} and \citet{whd2000})
and complications
arising from the uncertain and spatially varying instrument response
(\eg we believe the failure of the model to fit the spectrum of region 
NC in the energy range 1.6 -- 2.0 keV to be due to problems in the
spectral response used), rather than true evidence of low metal 
abundances in the gas. 

This spectral model was
used as a template fit to all the separate regions along the outflow 
lobes (see Fig.~3), 
allowing only absorption column density, gas temperature and 
absolute model normalization to vary (see Table.~1). 
Best fit hydrogen columns vary by a factor $\sim 45$, from
$\nH \approx (9\pm{1}) \times 10^{21} \pcmsq$ (Northern region N1)
to  $\nH \approx 2 \times 10^{20} \pcmsq$ (Southern region CLB). In contrast
characteristic temperatures vary by less than a factor of 2, from
$kT \approx 0.66^{+0.10}_{-0.08} \keV$ (NC) to 
$kT \approx 0.46^{+0.11}_{-0.10} \keV$ (S3).

\subsection{Hardness map}
\label{sec:hardness}

To fully exploit {\it Chandra}'s spectral-imaging capabilities
we constructed the ``hardness map,'' shown in Fig.~3, of the central
$\sim 1.5$ kiloparsec square region of NGC 253. This is a three color image
of the data between (approximately)
0.3 -- 0.6 keV, 0.6 -- 1.1 keV and 1.1 -- 2.0 keV in energy.
The southern outflow
cone is clearly much softer (a higher flux of low energy X-ray photons)
than the diffuse emission from the
starburst region, which itself is softer than the emission seen
from the northern outflow on the far side of the disk, consistent
with the simple spectral fitting discussed above.

These three energy bands were chosen based on the observed spectral
shape of the diffuse emission spectra. Emission in the 0.3 -- 0.6 keV energy
band, relatively prominent in the S1 to S3 and CLB regions,
is the first to be absorbed away as the foreground absorption column
increases. The relative prominence of the 0.6 -- 1.1 keV hump
(which contains the unresolved Fe L shell complex of lines)
in the nuclear region (region NC) is largely due to the higher column in this
region having removed much of the emission below E = 0.6 keV.
In region N1 the column is high enough to absorb away much of
emission between 0.6 -- 1.1 keV, making the 1.1 -- 2.0 keV energy band
appear much stronger (and hence the spectrum harder)
 than in the other spectra. As the spectral
fits in Table~1 show, the variation in spectral hardness of the
diffuse emission is mainly due to increasing absorption.

Note that the nucleus of the galaxy lies near the north western
edge of region NC (compare Figs~1 \& 3), so that region N1
partially covers the most obscured part of the starburst region
as well as the innermost part of the northern outflow lobe. 

The measured in-flight spectral resolution of the ACIS-S3 chip is
$\sim 0.12$ keV (full width at half maximum), at the operating
temperature of -110 $\degr C$ that applied to these observations 
(see the {\it Chandra} Observatory Guide). The three energy bands
used in Fig.~3 are therefore largely independent of each other (as
can be seen from the spectra),
\ie ``cross-talk'' between the bands is minimal.

These {\it Chandra} observations show that emission from the wind is spectrally
complex, due to spatial variations in absorption over scales as small
as $\sim 5\arcsec$ ($\sim 60\pc$), 
possibly with multiple temperature components
on even smaller scales. The southern outflow cone is relatively
spectrally uniform,
having risen above the location of absorbing gas in NGC 253's disk.
Without {\em high spatial resolution}, X-ray spectroscopy of diffuse
X-ray emission in star-forming galaxies will continue to yield
ambiguous results due to mixing regions of different temperature 
and/or absorbing column together, in addition to problems due to
 blending unrelated  point sources into the spectra.

\bigskip
\begin{minipage}[l]{0.46\textwidth}
\centerline{
\psfig{figure=strickland.fig4.ps,%
width={\columnwidth},angle=-90
}
}
  \medskip
  \noindent\begin{minipage}{0.99\linewidth} 
    \vskip \abovecaptionskip\footnotesize\noindent
  Fig. 4. --- {\it Chandra} ACIS S3 chip 
  X-ray spectrum of the limb brightened
  region of the southen outflow lobe (region CLB), fitted with
  a single temperature {\sc Mekal} hot plasma model. Note the
  lack of any additional hard X-ray component at energies
  above $2$ keV. See \S~\ref{sec:spectra}
  and Table~1 for details of the spectral fit.
   \par\vskip \belowcaptionskip
  \end{minipage}
\end{minipage}


\section{Spatial structure of the outflow}
\label{sec:outflow_structure}

\setcounter{figure}{4}

\begin{figure*}[!th]
 \centerline{
 \psfig{figure=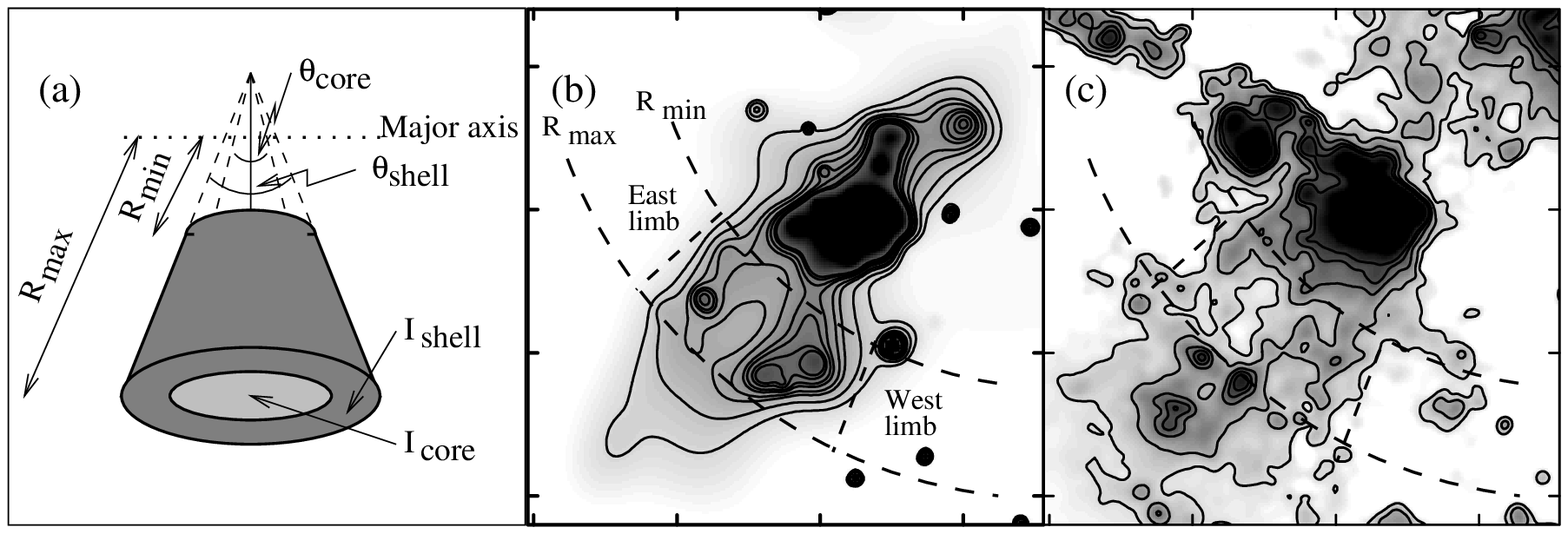,%
width={2\columnwidth}
}
}
 \caption{(a) Spatial model assumed for the origin of the X-ray and
   \halpha~emission. (b) {\it Chandra} $0.3$ -- $2.0 \keV$ image, with
   the region of clear limb-brightening between $R_{\rm min}$ and 
   $R_{\rm max}$, used in extracting 
   the azimuthal surface brightness profile, shown by the curved dashed lines.
   This region lies between 
   $27\arcsec$ and $47\arcsec$ ($\sim 340$ -- $590 \pc$)
   from the major axis of NGC 253. These lines, along with the dashed lines 
   parallel to the outflow, delimit the spatial region used to calculate
   gas masses, densities and filling factors as 
   described in \S~\ref{sec:outflow_structure}.
   (c) As (b), but showing the same region on the \halpha~image.
  }
 \label{fig:n253_fig_geom}
\end{figure*}

The striking similarity between X-ray and \halpha~emission in the southern 
outflow cone can be used to constrain the origin of the soft X-ray
emission seen in the superwind, and its physical properties. 
Although it had been known from even early {\it Einstein} X-ray observations
that the general spatial extent and distribution of X-ray-emitting
gas in NGC 253 and M82 was similar  to the \halpha~emission 
(\citet{watson84}; \citet{fab88})
the $\sim 5\arcsec$ spatial resolution, 
the $3$ -- $5\arcsec$ positional uncertainties, 
and limited sensitivity, of both  {\it Einstein}
\& {\it ROSAT} High Resolution Imagers (HRIs) only implied
that X-ray and \halpha~emission were similar over scales 
$\gtrsim 60$ -- $100 \pc$. In contrast, our {\it Chandra} observations 
can be used to investigate the X-ray / \halpha~correlation 
down to the effective resolution of the data ($\sim 0\farcs9 \approx 11 \pc$).

\subsection{Azimuthal surface brightness profiles}
\label{sec:outflow_profiles}

We looked at both azimuthal surface brightness profiles of
the limb-brightened region, and at
surface brightness ``slices'' constructed perpendicular to the
minor axis of the galaxy. Both methods give similar results, but 
azimuthal profiles have the advantage of preserving the conical,
diverging, structure of the outflow. 
We use the degree of limb-brightening
to constrain the fraction of X-ray and \halpha~emission that arises
in the shell or wall of the outflow, compared to emission from within the
central core of the outflow itself.

We created azimuthal profiles of background-subtracted 
X-ray and \halpha~surface brightness,
from the region of clear limb-brightening (see Fig.~\ref{fig:n253_fig_geom}),
using $1.0\degr$ and $0.5\degr$ azimuthal bins respectively, taking
care to exclude regions of emission from unrelated point sources.
The resulting surface brightness profiles are shown in 
Fig.~6.

Both X-ray and \halpha~azimuthal profiles peak at the same position
angle on the western limb of the outflow. In both cases the
eastern limb of the outflow cone is $\sim 0.75$ the intensity of the
western limb, and the surface brightness decreases inwards of the
eastern limb to a minimum value of $\sim 0.3$ -- $0.4$ of
the peak brightness (although the X-ray peak is slightly offset
to the interior of the cone from the \halpha).
There is not such a clear decrease in
brightness inwards of the western limb,
but the images (Fig.~\ref{fig:n253_fig_geom})
show additional ``clouds'' apparently within the
outflow in both X-ray and \halpha~images. These features might even be
unrelated emission associated with the underlying disk, 
rather than part of the outflow itself.

We use a simple model to quantify the fraction of emission
coming from within the core of the outflow, and what fraction comes from
the apparent shell. 
We assume a conical outflow, with a uniform emissivity core $I_{\rm core}$ of
opening angle $\theta_{\rm core}$, surrounded by a shell of emissivity
$I_{\rm shell}$ and outer opening angle $\theta_{\rm shell}$
(see Fig.~\ref{fig:n253_fig_geom}).
Predicted azimuthal profiles are convolved with Gaussian mask,
to take account for the effective spatial resolution of the X-ray and 
\halpha~data (FWHM = $0\farcs9$ [$\equiv0.6\degr$ in position angle
in the azimuthal profiles] for the X-ray and $1\farcs5$ for 
the \halpha~data), and compared by eye with the data. We primarily
attempt to match the position, width and intensity of the eastern limb,
given the more confused state of the western side of the profiles.
The relative intensity of the eastern limb with respect to the central
surface brightness minimum then constrains the emissivity of any
material within the core of the outflow. The best fit-by-eye shell
models are shown in Fig.~6. 
Attempting to
better match the intensity of the western limb would increase the peak-to-core
intensity contrast and lead to lower values of $I_{\rm core}/I_{\rm shell}$.

Both the X-ray and the \halpha~data are consistent with a hollow
($I_{\rm core}/I_{\rm shell}$ = 0) conical outflow with
$\theta_{\rm shell} = 26\degr$ and $\theta_{\rm core} = 20\degr$. 
The maximum value
of $I_{\rm core}/I_{\rm shell}$ consistent with the data
is $\sim 0.2$ (X-ray) or $\sim 0.1$ (\halpha). Both the X-ray
and \halpha~emission must therefore arise predominantly 
in a thin boundary layer,
between a faster and/or hot tenuous wind of SN-ejecta (that invisibly
fills the core of the outflow), and dense
cool ambient ISM surrounding the walls of the cavity.

We looked for spectral variation between the center and walls 
of the southern outflow cone, in an effort to detect 
hotter gas from within the cone, but found no statistically
significant differences.
We divided the limb brightened region into three sectors,
corresponding to the eastern limb, the low surface brightness core,
and the western limb, based on the azimuthal profile
shown in Fig.~6. 
As there were too few counts
in these sub-regions for spectral fitting we constructed
a hardness ratio $Q$ from the background subtracted count rates in the 
two energy bands
0.3 -- 1.1 keV and 1.1 -- 2.0 keV, and
calculated $Q$ for the entire limb brightened region, as well
as the three sub-regions (see Table.~2). Simulations with
{\sc Xspec} had shown this choice of energy bands to 
be reasonably sensitive to changes
in spectral hardness for thermal emission with temperature $kT \sim 0.1$ to
$3.0$ keV and hydrogen columns of between a few $\times 10^{20}$ to
a few $\times 10^{21} \pcmsq$. This hardness ratio
is more sensitive to temperature variation
than using a simple ratio of 0.3 -- 2.0 keV 
verses 2.0 - 8.0 keV count rates, for example.
Both the limbs and core of the limb brightened region have hardness
ratios statistically indistinguishable from one another or from the
region as a whole.
This is consistent with the majority of the emission, even within the
projected center of the outflow cone, being from the bright walls
of the cavity. Nevertheless, the uncertainty in the hardness ratio
of the core region is large enough to correspond to an uncertainty in
temperature (from the nominal $kT \sim 0.6$ keV) of approximately
a factor two.

The spectral uniformity of the southern outflow cone allows us to
interpret variations in surface brightness directly as variations
in emission integral (EI = $\int n_{\rm e} n_{\rm H} dV$).
Attempting to explain the central surface-brightness decrement of
$\sim 50$ -- $60$\%, with respect to the apparent walls of the cone,
as being purely due to absorption would require an order-of-magnitude
increase in absorption column in front of the center of the cone. 
This would also make the surprising degree of similarity
between the \halpha~and X-ray azimuthal profiles an astounding
coincidence, which seems unlikely.
For example, assuming emission from a hot plasma with
parameters equal to the best-fit model for the CLB region
(see Table.~1)
for the ``walls'' of the cavity, then reducing the flux
in the center of the cone to 40\% of its original value by extra
absorption requires increasing the column by a factor 12 to 
$\nH \sim 2.5 \times 10^{21} \pcmsq$. This would lead to
a strong decrement in the ratio of 0.3 -- 0.6 keV to 0.6 -- 1.1 keV
count rates in the center of the cone, 
relative to the walls of the cone, which is not seen.
Constructing a hardness ratio from these two energy bands, in a manner
analogous to those shown in Table~2, gives a
hardness ratio of $0.30\pm{0.12}$ for the center of the cone, 
compared to a value of $0.35\pm{0.04}$ for the entire CLB region
and the predicted value of $0.73$ for $\nH = 2.5 \times 10^{21} \pcmsq$.

The inferred opening angle of the outflow, $\theta \sim 26\degr$, is
very similar to the $\theta \sim 30\degr$ opening angle for the
outflow in M82. The inferred diameter of base of the outflow, \ie 
in the starburst region, is $\sim 24\arcsec$ ($300\pc$) which agrees very well
with the $\sim20\arcsec$ diameter region over which young SNRs are seen
in radio observations \citep{ua97},
consistent with the idea of a SN-driven outflow.

\subsection{Derived physical properties of the outflow}
\label{sec:outflow_model}

The maximum volume filling factor of the X-ray \& \halpha~emitting
material in the outflow cone is then $\eta \sim 0.4$, based
on the assumed geometry. If both X-ray \& \halpha-emitting gas
arise in a shocked or conductively heated boundary layer, 
in close proximity to each other, then their pressures should be similar
on {\em local scales} (\ie $\lesssim$ a few $100$ pc. On larger
scales the outflow is driven by the over-pressure of the starburst
region compared to the ambient ISM or IGM).
We can then infer the true density of the X-ray emitting gas
and its filling factor. In a $30\degr$-wide azimuthal region
centered on the position angle of the outflow cone and between
the radii defined in Fig.~\ref{fig:n253_fig_geom}, the
emission integral $EI \approx 5.5 \times 10^{60} \pcc$.
This is based on the observed $0.3$ -- $2.0 \keV$ {\it Chandra} 
ACIS-S count rate of $0.070 \pm{0.002}$ counts
$\ps$in this region, and assuming for simplicity
emission from a $kT = 0.5 \keV$ hot
plasma of Solar abundance, absorbed only by the Galactic foreground
hydrogen column of $\nH = 10^{20} \pcmsq$. The total volume (\ie
shell + core) of the conical frustum defined by our assumed geometry
is $V = 1.4 \times 10^{63} \cm^{3}$, which then gives
 $n_{\rm e} \approx 6.2 \times 10^{-2} \eta^{-1/2} \pcc$.
From \citet{mhvb87}, 
$P_{H\alpha}/k \sim 1$ to 
$4 \times 10^{6} \K \pcc$ in the \halpha~emitting gas, in the
region between $250$ and $500 \pc$ from the nucleus.
The true density and volume filling factor of the X-ray-emitting
gas is then $n_{\rm e} \sim 0.1$ to $0.3 \pcc$ and $\eta \sim 40$ to $4$
per cent (equivalently  $\sim 100$ to $10$ per cent of the geometrical 
volume of the shell).

The main  uncertainty affecting the derived EI is the assumed metal abundance.
As we believe the fitted metal abundance to be unrealistically low,
for the reasons given in \S~\ref{sec:spectra}, 
we prefer to assume a plausible
model for the emission instead of relying too heavily on
simplistic spectral fits. Assuming the gas temperature is 
$kT \sim 0.6$ keV instead of 0.5 keV only alters the EI by $\sim0.5$ per cent.

\bigskip
\begin{minipage}[l]{0.46\textwidth}
\centerline{
\psfig{figure=strickland.fig6.eps,%
width={\columnwidth},bbllx=70bp,bblly=42bp,bburx=435bp,bbury=675bp,clip=t}%
}
  \medskip
  \noindent\begin{minipage}{0.99\linewidth} 
    \vskip \abovecaptionskip\footnotesize\noindent
    Fig. 6. --- (a) Normalized azimuthal surface brightness 
    profiles of the southern outflow
    cone, in \halpha~emission (diamonds, dashed line) and $0.3$ -- 
    $2.0 \keV$ X-ray emission (crosses, dotted line). 
   (b) \halpha~profile, with best-fit hollow cone model 
   ($I_{\rm core}/I_{\rm shell} = 0$) overlaid 
   (solid line). A model with $I_{\rm core}/I_{\rm shell} = 0.2$ 
   (dashed line) over-predicts the central surface brightness
   between position angles $134$ -- $142\degr$. 
   (c) As (b), but for the soft X-ray emission.  Models with 
    $I_{\rm core}/I_{\rm shell} \ge 0.3$ (dashed line) fail to
   fit the data between position angles $134$ -- $142\degr$.
   \par\vskip \belowcaptionskip
  \end{minipage}
\end{minipage}

The total mass in X-ray emitting gas, within this specific region,
is $M_{\rm X-ray} \sim 1.5$ -- $6 \times 10^{4} \Msol$.
We used the spectroscopic observations of \citet{keel84} 
to flux calibrate our
\halpha~image (assuming a constant \halpha/\nii~ratio), and
hence calculate the \halpha~flux and gas mass in the same conical frustum
as used in the X-ray analysis.
Assuming the temperature of the \halpha~emitting gas is
$T_{\rm H\alpha} = 10^{4} \K$, and that the electron number density
of this gas is in the range $n_{e} = 100$ -- $25 \pcc$
(based on \citet{mhvb87}), 
the  mass of the \halpha~emitting gas in this region is
very similar to the mass of X-ray emitting gas, 
$M_{\rm H\alpha} \sim 1$ -- $5 \times 10^{4} \Msol$.
X-ray and \halpha~gas masses are similar, despite the large difference
in gas density, as the derived filling factor of the \halpha~emitting
gas within this region is substantially smaller than the filling
factor of the X-ray gas
($\eta_{\rm H\alpha} \sim 0.4$ -- $7 \times 10^{-4}$).

Low X-ray filling factors are  a prediction of current hydrodynamical
models of superwinds. In a 
parameter study of superwind models \citep{ss2000}, 
based roughly on M82's superwind, the
{\em globally-averaged}
 filling factor of the X-ray emitting gas in the {\it ROSAT} 
$0.1$ -- $2.4 \keV$ energy band (roughly equivalent to the
$0.3$ -- $2.0 \keV$ ACIS-S energy band we have used) was always
$\lesssim 2$ per cent.

Is it possible to have a wind powerful enough to compress
the ambient ISM to pressures $P/k \ge 10^{6} \K \pcc$, yet also tenuous
enough to be largely invisible in X-ray emission? In short, the answer is
yes!

The maximum emission integral of X-ray-emitting gas {\em not} from the
shell, based on $I_{\rm core}/I_{\rm shell} = 0.2$, is 
$EI_{\rm core} = 
1.2 \times 10^{60} \pcc$ ($kT = 0.5 \keV$) or $3.7 \times 10^{60} \pcc$ 
($kT = 2.0 \keV$, a realistic value for the
wind fluid temperature in this region based on the CC model).
These numbers correspond to $\sim 0.015$ ACIS S3 count s$^{-1}$ in
the 0.3 -- 2.0 energy band, and imply count rates in the
hard 2.0 -- 8.0 keV energy band of $\sim 3 \times 10^{-4}$ (assuming $kT = 0.5
\keV$) or $1.6 \times 10^{-3}$ count s$^{-1}$ ($kT = 2.0$ keV).
The total observed 2.0 -- 8.0 keV count rate in this region
is only $(0.9\pm{1.6}) \times 10^{-3}$ count s$^{-1}$, which
can be accounted for purely by the $kT = 0.6$ keV emission from the
shell (extrapolation from the 0.3 -- 2.0 keV count rate
predicts a 2.0 -- 8.0 keV count rate of $\sim 1.4 \times 10^{-3}$ count
s$^{-1}$). This, and the hardness ratios given in Table~2, do not rule
out a spectrally harder core component to the X-ray emission
in the limb brightened region, given the poor statistics.
Nevertheless they suggest that the upper limits on the core
emission integrals given above are conservative.

In the CC model of a freely-expanding wind, ram pressure dominates over
thermal pressure at distances beyond a few starburst-region radii. For
our model to work we require a wind ram pressure such that
$P_{\rm ram } = \rho_{\rm w} v_{\rm w}^{2} \ge 
P_{\rm X-ray} \sim P_{\rm H\alpha}$, as it is the
wind that shocks and compresses the ambient ISM. This is a lower
limit --- the wind ram pressure $P_{\rm ram}$ can be several times
$P_{\rm H\alpha}$ as the $H\alpha$~emitting material is also being
swept-up and accelerated to velocities of several hundred kilometers
per second by the wind. Taking this into account, 
let us assume $P_{\rm ram} = \psi P_{\rm H\alpha}$.
Inspection of our existing hydrodynamical simulations of superwinds
shows $1 \le \psi \le 10$, there being a large scatter in $\psi$
from region to region within the wind. 
The {\em lower limit} on the emission integral of
the wind fluid, required to give acceptable wind ram pressures
in the same region as the limb-brightened \halpha~and
X-ray emission, is $EI_{\rm w} \gtrsim V 
(1 - \eta_{\rm X-ray} - \eta_{\rm H\alpha}) \times 
(\psi P_{\rm H\alpha}/2 \mu m_{\rm H} v_{\rm w}^{2})^{2} 
\sim 1$ -- $13 \times 10^{57} \psi^{2} v_{3}^{-4} \pcc$, 
where $v_{3}$ is the wind velocity in units of $3000 \kmps$ 
($v_{3} \sim 1$ for standard
mass and energy injection rates, \eg see \citet{lh95}),
and assuming $\psi = 1$ to give a lower limit.
Assuming that $\psi = 3$ is typical of the wind on average,
then the wind velocity must be $v_{3} \ge 0.3$ -- $0.6$, in order not
to violate the observed upper limits on wind emission integral
(using the lower, and more conservative,
 of the two values given in the previous paragraph),
which seems entirely reasonable.

These limits also make sense in terms of mass transport rates in the
wind. The observed SN rate of $\sim 0.05 \pyr$ 
\citep{colina92} 
corresponds to an expected $\Mdot_{\rm w} \sim 0.5 \Msol \pyr$,
using the \citet{lh95} models, and assuming no additional mass-loading.
Based on the pressure balance argument, the total mass
loss rate (assuming two identical outflow cones) in the wind fluid alone
is $\Mdot_{\rm w} \gtrsim 0.032$ -- $0.13 \, \psi v_{3}^{-1} \Msol \pyr$.
The observational upper limit on the emission 
integral of the core of the outflow gives $\Mdot_{\rm w} < 2.2 
v_{3} \Msol \pyr$, using the higher of the two $EI_{\rm core}$ estimates.

It is interesting to note that a clear detection of the
wind component would have placed strong constraints on wind density, velocity,
energetics and mass outflow rate. Separating the faint X-ray emission
due to the wind from the brighter X-ray emission from
wind/ISM interfaces in local starburst galaxies
requires arcsecond spatial resolution --- a job
only {\it Chandra} can do.


\subsection{Additional discussion}
\label{sec:outflow_additional}

How robust are these conclusions on the origin and filling factor
of the X-ray emitting gas, given the limited size of the region of 
unambiguous limb-brightening, the simplicity of the spatial model
assumed and the uncertain value of the gas emission integral?

Although these factors probably make the estimated
densities and filling factors of the X-ray emitting gas uncertain to an
order of magnitude or so, the general picture of the X-ray and 
\halpha~emission being dominated by moderate-to-low filling factor 
gas, arising in the interaction between the wind and denser ambient ISM,
is likely to remain true.
The simple analytical arguments presented in \S~\ref{sec:outflow_model}
certainly demonstrate that our conclusions are physically plausible.

Within the region we have considered, the filling factor 
of the gas that is the 
dominant X-ray emitter {\em must} be less than unity, as it
is impossible for a volume-filling gas to reproduce the observed
limb-brightening without invoking fortuitous absorption along
the center of the outflow cone. Such additional
absorption would be detectable in the spectra and hardness
maps, and is therefore inconsistent with the observed spectral
uniformity of the southern outflow cone.

That strong limb-brightening is only observed in a narrow
range of heights above the plane of the galaxy is not
surprising. Within the immediate vicinity of the starburst
region \halpha~and X-ray emission not directly associated with
the superwind itself will hide any limb-brightened outflow
component. At much larger distances, as the surface
brightness of the superwind decreases, limb-brightening
would become increasingly difficult to detect. A well defined
cavity or chimney structure with dense walls might only exist
within a few scale heights of the disk. Further out in the
wind, the dense gas in the outflow probably
exists as isolated dense clumps and clouds dragged out of the disk
(as seen in hydrodynamical simulations, \eg \citet{ss2000}).
These dense clouds, although of low filling
factor and again dominating the X-ray and optical emission,
would not necessarily give a limb-brightened optical or X-ray morphology.

In principle, the filling factor of the X-ray emitting gas
might be different in other parts of the flow. If density
inhomogeneities in the flow are destroyed over a time scale
less than or equal to the flow time, then the X-ray emission
in the outer parts of the wind might be dominated by a homogenized
volume-filling hot component of the wind. Detecting filamentary
or ``clumpy'' X-ray emission at lower surface brightness
in the outer parts of the wind (with {\it XMM-Newton} or 
{\it Chandra}) would disprove this hypothesis.
However, even if the X-ray emission appears relatively uniform
the filling factor is not necessarily high. It is easy to 
conceive of situations where low filling factor clouds
dominate the X-ray emission but whose spatial distribution
is so uniform as to appear smooth.

It is only with the spatial resolution {\it Chandra} provides
that it is possible to resolve (both spectrally and spatially)
the inner outflow cones (within
a few hundred parsecs of the starburst region) where
unambiguous evidence of the origin of the soft X-ray
emission in superwinds can be found.

\section{Summary}
\label{sec:discussion}

{\it Chandra} ACIS observations demonstrate that 
the soft X-ray emission in the less-obscured
of NGC 253's superwind's nuclear outflow cones is well collimated, and
extremely similar in morphology and extent to the previously observed
limb-brightened \halpha~outflow cone. 
This provides the first direct observational evidence that both 
X-ray and \halpha~emission come from low volume filling factor gas,
regions of interaction  between a tenuous starburst-driven 
wind of SN-ejecta and the dense ambient ISM, and not from the wind itself.

We measure the observationally
and theoretically important filling factor of the
X-ray emitting gas at between $\eta \sim 4$ and $40$ per cent. The
observed morphology of the outflow also provides a purely geometrical
upper limit of $\eta \sim 40$ per cent.
Only $\lesssim 20$ per cent of the observed X-ray emission
can come from the volume-filling, metal-enriched, wind fluid itself.
Spatially-resolved spectroscopy of the soft diffuse thermal X-ray emission
reveals that the predominant source of spectral variation along the
outflow cones is due to strong variation in the absorption on scales
of $\sim 60 \pc$, with little evidence for any change in the characteristic
temperature of the emission.
These observations are easily explained by, and fully consistent
with, the standard model of a superwind driven by a starburst of
NGC 253's observed properties. 

The important implication of this work is that we can not see the majority
the energy and gas (in particular the hot metal-enriched gas) transported
out of NGC 253 by it's superwind, even in X-ray emission.
If this is generally true for all starbursts, the majority of
IGM heating and enrichment by starbursts may be currently invisible in
{\em emission}. This emphasizes the need for, and importance of, absorption
line studies of warm and hot gas in these fascinating and important objects.

\acknowledgments

We would like to thank the anonymous referee for their constructive
criticism. The team that built and operates {\it Chandra} also deserves
recognition, for their hard work, and for making such an excellent
telescope.
This work was supported by NASA through grants LTSA NAG56400 and GO0-1008X.



\end{document}